\documentclass[prl,aps,amsmath,twocolumn,showpacs,superscriptaddress]{revtex4-1}

\usepackage{amsmath}
\usepackage{mathrsfs}
\usepackage{amsfonts}
\usepackage{graphicx}
\usepackage{color}
\begin{document}

\title{Experimental validation of the theoretical prediction for the optical 
$S$ matrix}

\author{A.~M. Mart\'inez-Arg\"uello}
%\email{blitzkriegheinkel@gmail.com}
\affiliation{Instituto de Ciencias F\'isicas, Universidad Nacional Aut\'onoma 
de M\'exico, Apartado Postal 48-3, 62210 Cuernavaca, Mor., Mexico}

\author{V. Dom\'inguez-Rocha}
%\email{vidomr@gmail.com}
\affiliation{Instituto de Ciencias F\'isicas, Universidad Nacional Aut\'onoma 
de M\'exico, Apartado Postal 48-3, 62210 Cuernavaca, Mor., Mexico}

\author{R.~A. M\'endez-S\'anchez}
%\email{mendez@icf.unam.mx}
\affiliation{Instituto de Ciencias F\'isicas, Universidad Nacional Aut\'onoma 
de M\'exico, Apartado Postal 48-3, 62210 Cuernavaca, Mor., Mexico}

\author{M. Mart\'inez-Mares}
%\email{moi@xanum.uam.mx}
\affiliation{Departamento de F\'isica, Universidad Aut\'onoma
Metropolitana-Iztapalapa, Apartado Postal 55-534, 09340 Ciudad de M\'exico, 
Mexico}

\begin{abstract}
Scattering of waves is omnipresent in nature in systems with sizes varying from $10^{-15}$ to $10^{25}$~m. 
Within this 40 orders of magnitude, in a great number of systems, the scattering can be separated in an averaged response that crosses rapidly the scattering region and a fluctuating delayed response. 
This fact is the basis of the optical model; the averaged response, represented by the optical matrix $\langle S\rangle$, is composed with the fluctuating part that can be taken as a random matrix. 
Although the optical model was developed more than 60 years ago, a theoretical prediction for the optical matrix was obtained until very recently. 
The validity of such prediction is experimentally demonstrated here. 
This is done studying the scattering of torsional waves in a quasi-1D elastic system in which a locally periodic system is built; the distribution of the scattering matrix is calculated completely free of parameters. 
In contradistinction to all previous works, in microwaves and in elasticity, in which the value of $\langle S\rangle$ is obtained from the experiment, here the theoretical prediction is used to compare with the experiment. 
Numerical simulations show that the theoretical value is still valid when strong disorder is present. 
Several applications of the theoretical expression for the optical matrix in other areas of physics are proposed. 
Possible extensions of this work are also discussed.
\end{abstract}
 

\maketitle

The scattering amplitudes, in almost all physical wave systems, can be 
separated in a component passing rapidly through the scattering region plus a 
delayed response coming from multiple scattering. This fact is summarized in the 
optical model in which the dispersion amplitudes are separated into an averaged 
part and a fluctuating part. The averaged response is captured by the optical 
matrix $\langle S\rangle$ while the fluctuations are commonly studied using 
statistical techniques from random matrix theory. Therefore, the optical matrix 
has become a fundamental quantity in the description of multiple scattering of 
particles and waves. It was introduced in the optical model of the nucleus 
developed in the 1950's by Feshbach, Porter and 
Weisskopf~\cite{Feshbach,FeshbachPorterWeisskopf}. Since the scattering of a 
nucleon by an atomic nucleus is equivalent to the theory of 
waveguides~\cite{EricsonMayer-Kuckuk}, this model has been extended not only to 
chemical reactions but also to electronic transport through ballistic quantum 
dots and microwave cavities~\cite{MelloBaranger,MelloKumar} and more recently to 
mechanical waves~\cite{BaezEtAl,
Martinez-ArguelloMartinez-MaresCobian-SuarezBaezMendez-Sanchez}. 

The average $\langle S\rangle$ can be physically interpreted as the fraction 
of the incident wave packet which comes out promptly from the scattering 
region~\cite{Feshbach}. A concrete realization of $\langle S\rangle$ was 
proposed in Ref.~\cite{BrouwerBeenakker} in the transport of electrons in 
mesoscopic systems: when the incoming and outgoing channels are not coupled 
perfectly to the internal system, the optical matrix $\langle S\rangle$ 
quantifies the coupling between the internal and external regions. This is very 
important since the imperfect coupling has to be taken into account in almost 
all scattering experiments~\cite{BaezEtAl,Mendez-SanchezEtAl2003,Hemmadyetal,
Kuhletal,LawniczakBauchHulSirko,AureganPagneux,Martinez-ArguelloEtAl,LawniczakSirko}. 
Furthermore, as a consequence of the 
analyticity of the $S$ matrix, even when absorption is 
present~\cite{GoparMartinez-MaresMendez-Sanchez,
Martinez-ArguelloMendez-SanchezMartinez-Mares}, wave scattering systems are 
self-averaging~\cite{MelloKumar}. This means that $\langle S\rangle$ is the 
only relevant parameter needed to obtain all scattering properties in complex 
systems since the fluctuations seem to be universal depending only on very 
general symmetry properties as presence/absence of time reversal invariance, 
among others. 

There are, on the one hand, several theoretical studies in which 
$\langle S\rangle$ is used to obtain the distribution of the scattering matrix 
known as Poisson's 
kernel~\cite{LopezMelloSeligman,MelloPereyraSeligman,FriedmanMello,
FyodorovSavin}. In the one channel case it reads
\begin{equation}
\label{eq:KPoisson}
p_{\langle S\rangle}(S) = \frac{1}{2\pi} \,
\frac{1-|\langle S\rangle|^2}{|S-\langle S\rangle|^2},
\end{equation}
where $S=\mathrm{e}^{\mathrm{i}\theta}$. An analytical value for $\langle 
S\rangle$ was missing during more than 60 years; it was obtained only very 
recently for a one-dimensional chain of delta 
potentials~\cite{Dominguez-RochaMendez-SanchezMartinez-Mares}. On the other 
hand, up to now, there are no experimental studies about $\langle S\rangle$. 
In all experiments performed in microwave cavities and graphs, and in elastic 
systems, the  value of $\langle S\rangle$ is obtained afterwards from the 
measurement since an analytical expression for it was not available at that 
time. Then, to compare with the experiment, the numerical value obtained from 
the experiment is used in the theoretical expression of Poisson's kernel in a 
kind of self-consistent argument.

Since the experiment is the only mechanism to validate a theoretical development, in this paper, using elastic waves, we experimentally demonstrate the validity of the theoretical prediction for the optical matrix $\langle S\rangle$ given in 
Ref.~\cite{Dominguez-RochaMendez-SanchezMartinez-Mares}. 
To do this the scattering of torsional waves in a beam, in which a finite crystalline structure is machined, is studied. 
In contradistinction with the methods of Refs.~\cite{Griffiths,Luna-AcostaEtAl}, here, to obtain the optical matrix in this system, the scattering formalism developed in Ref.~\cite{Dominguez-RochaMartinez-Mares} is applied to the elastic crystalline structure. 
As we will see below the distribution of the scattering matrix, with 
the theoretical value of $\langle S \rangle$, correctly predicts the results measured with acoustic resonant spectroscopy completely free of parameters. 
Numerical results when disorder is present, are also given. 

\begin{figure}
\includegraphics[width=\columnwidth]{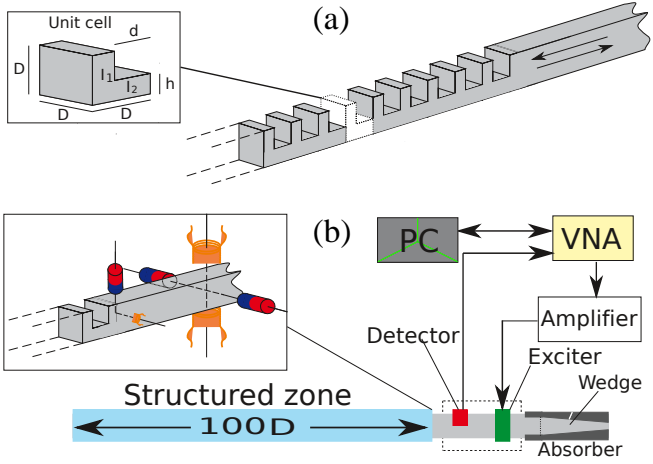}
\caption{(a) Aluminum beam ($\sqrt{G/\rho}=3104.7~\mathrm{m/s}$) which consists of a region with a locally periodic structure, composed of $N$ scatterers, and a semi-infinite uniform region of squared cross-section. The unit cell (inset) of length $D$ consists of bodies 1 and 2 of lengths $D-d$ and $d$, with polar moments of inertia $I_1$ and $I_2$, 
respectively. We use $w_1=h_1=w_2=D=2.54$~cm, $h_2=h=0.9525$~cm and $d=1.5875$~cm.
(b) Setup: a beam consists of a structured zone, a wedge zone with an absorber, and a free region where the exciter and detector are located. Inset: The coils of the EMAT are connected in series and their polarity is such that the EMAT excites only torsional waves. The EMAT detector is composed of a magnet and a small coil.}
\label{fig:Scheme}
\end{figure}

Lets consider the semi-infinite beam of Fig.~\ref{fig:Scheme} (a) in which a 
locally periodic structure of $N$ notches is machined, such that the unit cell 
is formed by two parts or bodies labeled as 1 and 2. Plane waves are 
sent to the structure from the uniform part and the response of the system is 
obtained using the scattering matrix formalism. The stationary solution 
$\psi_j(x)$  of the one-dimensional wave equation describing torsions, in the 
$j$th body of the unit cell of Fig.~\ref{fig:Scheme}, is a superposition of 
waves traveling to the left and to the right 
$\psi_j(x)=a_j\,\textrm{e}^{-\textrm{i}k_jx}+b_j\, \textrm{e}^{\textrm{i}k_jx}$,
where $j(= 1, 2)$ indicates the corresponding part of the unit cell and $k_j$
is the wave number related to the frequency by $k_j = 2\pi f/c_j $ with
$c_j=\sqrt{G\alpha_j/\rho I_j}$ the phase velocity of the torsional waves. 
Here, $\rho$ is the density of the rod, $G$ its shear modulus, $I_j$ the polar 
moment of inertia of part $j$, and $\alpha_j$ the Navier series,
\begin{equation}
\alpha_j = \sum_{m=0}^{\infty} 
\sum_{p=0}^{\infty} \frac{256/\pi^6}{(2m+1)^2(2p+1)^2 }
\frac{h_jw_j}
{\left(\frac{2m+1}{h_j}\right)^2+\left(\frac{2p+1}{w_j}\right)^2},
\end{equation}
with $w_j$ and $h_j$ the width and height of body $j$.

The boundary conditions between bodies 1 and 2, which are in contact at a point 
$x=x'$, are continuity of the wave amplitude, $\psi_1(x')=\psi_2(x')$, and 
continuity of the moment of torsion, $M_{T_1}(x')=M_{T_2}(x')$. The latter 
is related to the derivative of the wave amplitude through 
$M_{T_j}(x)=G\alpha_j\partial\psi_j(x)/\partial x$. Thus, 
the derivative of the wave amplitude, at the point $x=x'$, is discontinuous by a 
factor $\eta=\alpha_2/\alpha_1$ when the same material is used in bodies 1 and 
2. Since the two parts of the unit cell oscillate with the same frequency, the 
wave numbers of both bodies are related through 
$k_1=k_2\,c_2/c_1=k_2\sqrt{\eta I_1/I_2}$, where $I_1=\frac16 D^4$ and 
$I_2=\frac16 D^4\left(2\frac{h}{D}-3\frac{h^2}{D^2} + 2\frac{h^3}{D^3}\right)$ 
are the polar moments of inertia of bodies 1 and 2, of the unit cell of length 
$D$, respectively. 
To calculate $I_2$ the parallel axes theorem was used with a distance 
$(D-h)/2$.

\begin{figure}
\centering
\includegraphics[width=\columnwidth]{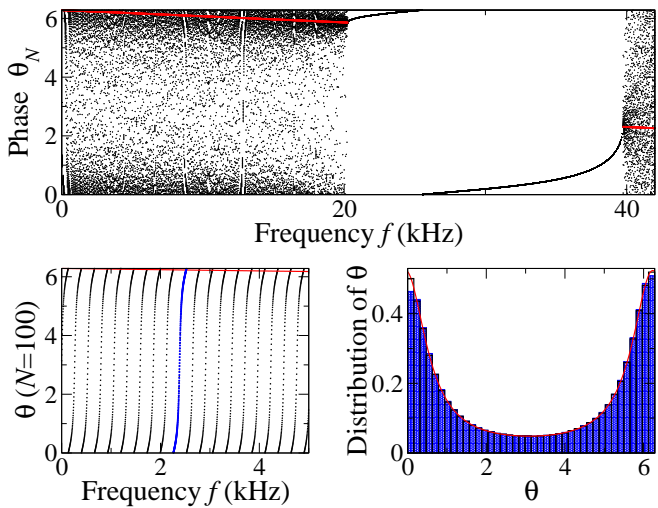}
\caption{(Color online) Upper panel: the last 15 iterations, of 1000, of Eq.~(\ref{eq:recurrence}), are plotted as a function of the frequency with an initial condition $\theta_0=\pi$. 
The red line corresponds to the phase of $\langle S\rangle$ in Eq.~(\ref{eq:Sav}). 
Lower panels: in the left panel the resonances of the phase of Eq.~(\ref{eq:recurrence}), for $N=100$, are shown. 
The histogram of the resonance highlighted in blue is shown in the right panel. 
More details of the right panel are given in the text.}
\label{fig:map}
\end{figure}

The reflection and transmission amplitudes through a single scatterer 
are
\begin{equation}
\label{eq:rntn}
r_\textrm{n} = \frac{ 2\textrm{i}\beta\, \sin(k_2 d) } 
{\beta^2\textrm{e}^{\textrm{i}k_2 d} -\textrm{e}^{-\textrm{i}k_2 d} }
\quad\mbox{and}\quad
t_\textrm{n} = \frac{ \beta^2 - 1}
{\beta^2 \textrm{e}^{\textrm{i}k_2 d} -\textrm{e}^{-\textrm{i}k_2 d}},
\end{equation}
where $\beta=(k_1-\eta k_2)/(k_1+\eta k_2)$ and $d$ is the scatterer length. 
These were obtained using the boundary conditions at both sides of the notch, 
solving the resulting system of equations.
%
%Now $\langle S\rangle$ will be obtained. 
The response of the system of $N$ 
scatterers is described by the $1\times 1$ scattering matrix $S_N$, which is 
related to the scattering matrix $S_{N-1}$, that describes the system with 
$N-1$ scatterers, through the following recurrence 
relation~\cite{Dominguez-RochaMartinez-Mares}
\begin{equation}
\label{eq:recurrence}
S_N = %\frac
{\left( r_\textrm{n}z^*_\textrm{n} + z_\textrm{n}S_{N-1} \right)} 
{\left(r^*_\textrm{n}z_\textrm{n} + z^*_\textrm{n}S^*_{N-1} \right)^{-1}} S^*_{N-1}
\end{equation}
where $z_\textrm{n}=t_\textrm{n}\textrm{e}^{\textrm{i}k_1(D-d)}$. The wave 
number $k_1$ is the tunable parameter, proportional to the frequency, since the 
uniform part of the beam has the same transversal area as body 1 of the unit 
cell. 
According to Refs.~\cite{Dominguez-RochaMartinez-Mares,
Dominguez-RochaMendez-SanchezMartinez-Mares,Martinez-MaresRobledo,VidomEPJST}, 
the recurrence relation given in Eq.~(\ref{eq:recurrence}) can be 
interpreted as a non-linear map, for the phase of the scattering matrix, that reveals the forbidden and allowed bands as a function of $k_1$ (see upper panel of 
Fig.~\ref{fig:map}). This non-linear maping accepts stable and unstable fixed point 
solutions when $N\to\infty$, the latter being a set of zero measure that will 
be ignored. A fixed point solution can be interpreted as the optical 
matrix, $\langle S\rangle$, since it satisfies the analyticity condition 
$\langle S^N\rangle=\langle 
S\rangle^N$~\cite{Dominguez-RochaMendez-SanchezMartinez-Mares}. The values of 
the optical matrix in an allowed band is (we are interested in this region 
only)
\begin{equation}
\label{eq:Sav}
\langle S\rangle = 
{\mathrm{i}}{(r^*_\textrm{n} z_\textrm{n})^{-1}}
\left[-\sqrt{|t_\textrm{n}|^4-(\mathrm{Re}\, 
z_\textrm{n})^2}+\mathrm{Im}\,z_\textrm{n} \right].
\end{equation}
It is remarkable that the optical matrix $\langle S\rangle$, and Poisson's 
kernel, Eq.~(\ref{eq:KPoisson}), is completely determined (and only depends) on 
the values of the reflection and transmission amplitudes of a single scatterer, 
no matter how they are obtained, by numerical or experimental methods or by a 
theoretical model as in Eq.~(\ref{eq:rntn}).

In the upper panel of Fig.~\ref{fig:map} the band structure in frequency is observed for the last 15 iterations of 1000. 
The allowed bands are clearly observed (the first band and a part of the second one are shown) since the iterations cover the full interval between 0 and $2\pi$ for the phase in a non uniform way. 
It can be also observed in the forbidden bands, only one is shown, 
that the iterations reach a single fixed point. 
The solution in the allowed bands, given by Eq.~(\ref{eq:Sav}), is also plotted. 
This solution corresponds to the maximum of the distribution around which $\theta_N$ is distributed. 
In the left lower panel of Fig.~\ref{fig:map} the phase $\theta_N$, for $N=100$, shows the resonances as a function of the frequency in an allowed band; resonances do not appear in the forbidden band. 
{As it can be seen in this figure, the phase do not increase linearly; the largest slope is associated to the resonant peaks.}
%once unwrapped, in Figure 2-4 with frequencies with the slope being largest around \theta =3. I guess it is related to the resonance peaks. Can the authors comment on this?
%
The distribution of the phase in the first allowed band is studied numerically in two different ways. 
Firstly, the distribution of the phase for $10^4$ 
realizations (iterations) for the fixed frequency $f=2395$~Hz, is plotted in 
the right lower panel of Fig.~\ref{fig:map} as a black histogram. Secondly, the 
(blue) bars in the same figure represent the histogram of the phase along the 
resonance centered at $f=2395$~Hz; {the maximum of the distribution corresponds to the lowest slope of the phase given in the left lower panel of the same figure.} This resonance is (blue) highlighted 
in the left lower panel. Also, in the same figure, the distribution of the 
Poisson kernel, given by Eq.~(\ref{eq:KPoisson}) with the average of the 
scattering matrix $\langle S\rangle$ of Eq.~(\ref{eq:Sav}) for $f=2395$~Hz, is 
plotted (continuous red curve). An excellent agreement between the numerical 
histograms and the theoretical distribution is obtained. This result is relevant 
for experimental implementation since it allows to study a specific resonance of a 
given sample, instead of several sample realizations. 

In what follows we will show that Poisson's kernel, with the average taken from Eq.~(\ref{eq:Sav}), correctly predicts the experimental distribution of the scattering matrix in elastic waves. 
This will be done for a system with a large, but fixed, number of scatterers within a small frequency range in the first allowed band. The experimental setup is shown in Fig.~\ref{fig:Scheme} (b). 
A signal of frequency $f$, produced by a vector network analyzer (VNA, Anritsu MS-4630B) and intensified by a Cerwin-Vega (CV-900) high-fidelity audio amplifier, is sent to an electromagnetic acoustic transducer (EMAT) designed {\em ad hoc} for this experiment since high power and selectivity is needed. 
This transducer, composed by two coils and two permanent magnets, shown in the 
inset, produces torsional vibrations that propagate through the 
system~\cite{MoralesFloresGutierrezMendez-Sanchez, 
Franco-VillafaneFlores-OlmedoBaezGandarilla-CarrilloMendez-Sanchez}. 
The response, measured by another EMAT, is directly sent to the VNA. The 
measurements, amplitude and phase, as a function of the frequency $f$, are taken 
from the VNA to the computer through a GPIB port.

The system under study consists of an aluminum beam of squared cross-section of 
width $D=2.54$~cm and length $3.6$~m divided in three regions. From a free 
boundary, a locally periodic structure, composed of 100 equal notches as in 
Fig.~\ref{fig:Scheme} (b), is machined. The middle  part of the 
beam, of 56~cm length, remains uniform. In this part the waves are excited and 
the scattering matrix is measured. The other end simulates a semi-infinite beam 
by means of a passive vibration isolation (PVI) system that absorbs the incoming 
waves~\cite{BaezEtAl,Martinez-ArguelloMartinez-MaresCobian-SuarezBaezMendez-Sanchez,Flores-OlmedoEtAl,Arreola-LucasEtAl}. 
The PVI system, is composed of a wedge and polymeric foams and has length 
of 50~cm, covering completely the wedge and part of the uniform section of the 
beam. This system %, designed and implemented only in our experiments, 
allows the measurement of the mechanical scattering matrix, in the frequency domain, since 
the normal modes of the complete beam cannot be established.

\begin{figure}
\centering
\includegraphics[width=\columnwidth]{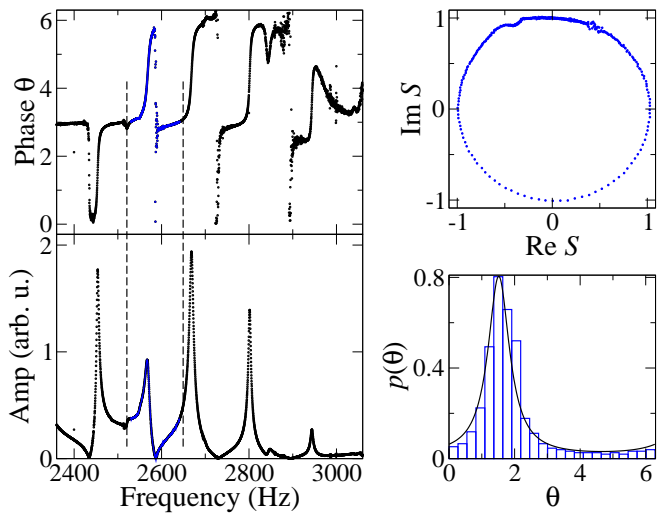}
\caption{(Color online) Amplitude and phase of the $S$-matrix as a function of 
the frequency (left panels). Shifted and normalized $S$-matrix in the Argand 
plane (upper right panel), in a part of the allowed band. The histogram of the 
highlighted resonance (blue) at $f=2570$~Hz and its comparison to the Poisson 
kernel, Eq.~(\ref{eq:KPoisson}) with $\langle S\rangle$ given by 
Eq.~(\ref{eq:Sav}), is shown in the right lower panel. The dimensions of the 
unit cell are the same as in Fig.~\ref{fig:Scheme} but $h$ replaced by an 
effective value, $0.746h$, due to the punching of body 
2~\cite{MoralesFloresGutierrezMendez-Sanchez}.}
\label{fig:Amp}
\end{figure}

In Fig.~\ref{fig:Amp} (left panels) the measured amplitude and phase of the 
scattering matrix, as a function of the frequency, for a part of the first band, 
is shown. As expected, several resonances of the allowed band, for which the 
phase takes values between 0 and $2\pi$, are observed. All of these resonances 
describe, in the Argand plane, non-concentric circles of different radii. This 
is due to the impedance of the detector~\cite{BaezEtAl,
Martinez-ArguelloMartinez-MaresCobian-SuarezBaezMendez-Sanchez}. 
We analyze the phase of the resonance lying between the dotted lines, from 
2530.0 to 2638.8~Hz, using the method of Ref.~\cite{BaezEtAl} to subtract 
the shift due to the impedance. As seen in the right upper panel of the same 
figure, this corrected $S$ matrix describes a circle centered at the origin (the 
radius was set to 1 for convenience). The distribution of the phase along the 
circle is shown in the lower panel as a histogram. In the same figure the 
analytical distribution expressed by the Poisson kernel, 
Eq.~(\ref{eq:KPoisson}), continuous line, is also given. The value of the 
optical matrix was taken from Eq.~(\ref{eq:Sav}) evaluated at $f=2570$~Hz, the 
center of the resonance. A very good agreement between theory and experiment 
is observed.
%Other frequency intervals were also analyzed and also show a very good agreement with the theoretical predictions, nothwistanding other factors as distance to the gap, etc. (See Fig.~\ref{fig:OtherResonance}).
The only effective parameter used is the one related to the punching of body 2 by body 1. This parameter appears because of the  one-dimensional character of the $S$-matrix theory used whereas the constructed beam, given in Fig.~\ref{fig:Scheme} (a), is three-dimensional.

\begin{figure}
\centering
\includegraphics[width=\columnwidth]{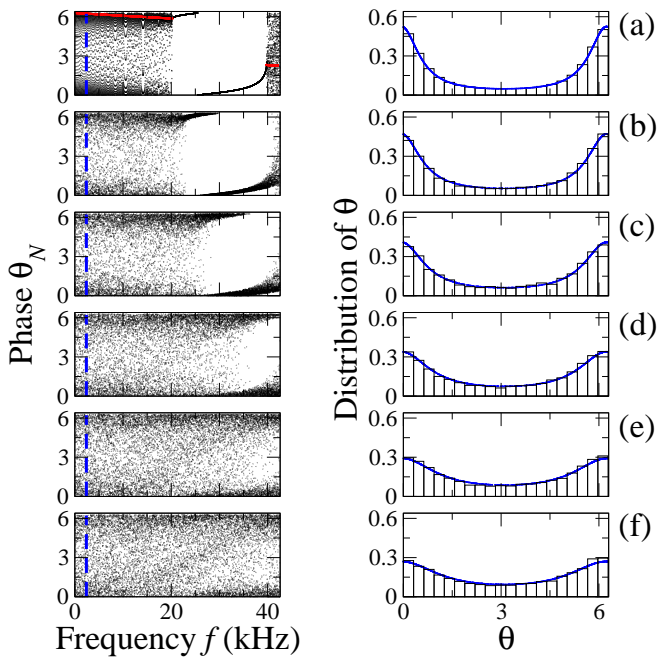}
\caption{(Color online) Left: the last 15 iterations, of 1000, of Eq.~(\ref{eq:recurrence}), are plotted as a function of the frequency with an initial condition $\theta_0=\pi$ for different degrees of disorder: (a) 0\%, (b) 10\%, (c) 25\%, (d) 50\%, (e) 75\% y (f) 90\%.
The red line corresponds to the phase of $\langle S\rangle$ in Eq.~(\ref{eq:Sav}); it is vallid for all panels at the left since only indistinguishable changes appear. 
The respective histograms of the phase are given in the right panels. 
The continuous curve is the Poisson kernel, Eq.~(\ref{eq:KPoisson}), with $\langle S\rangle$ given by Eq.~(\ref{eq:Sav}) averaged over the disorder realizations. 
}
\label{fig:disorder1}
\end{figure}

\begin{widetext}
\begin{figure*}
\centering
\includegraphics[width=2\columnwidth]{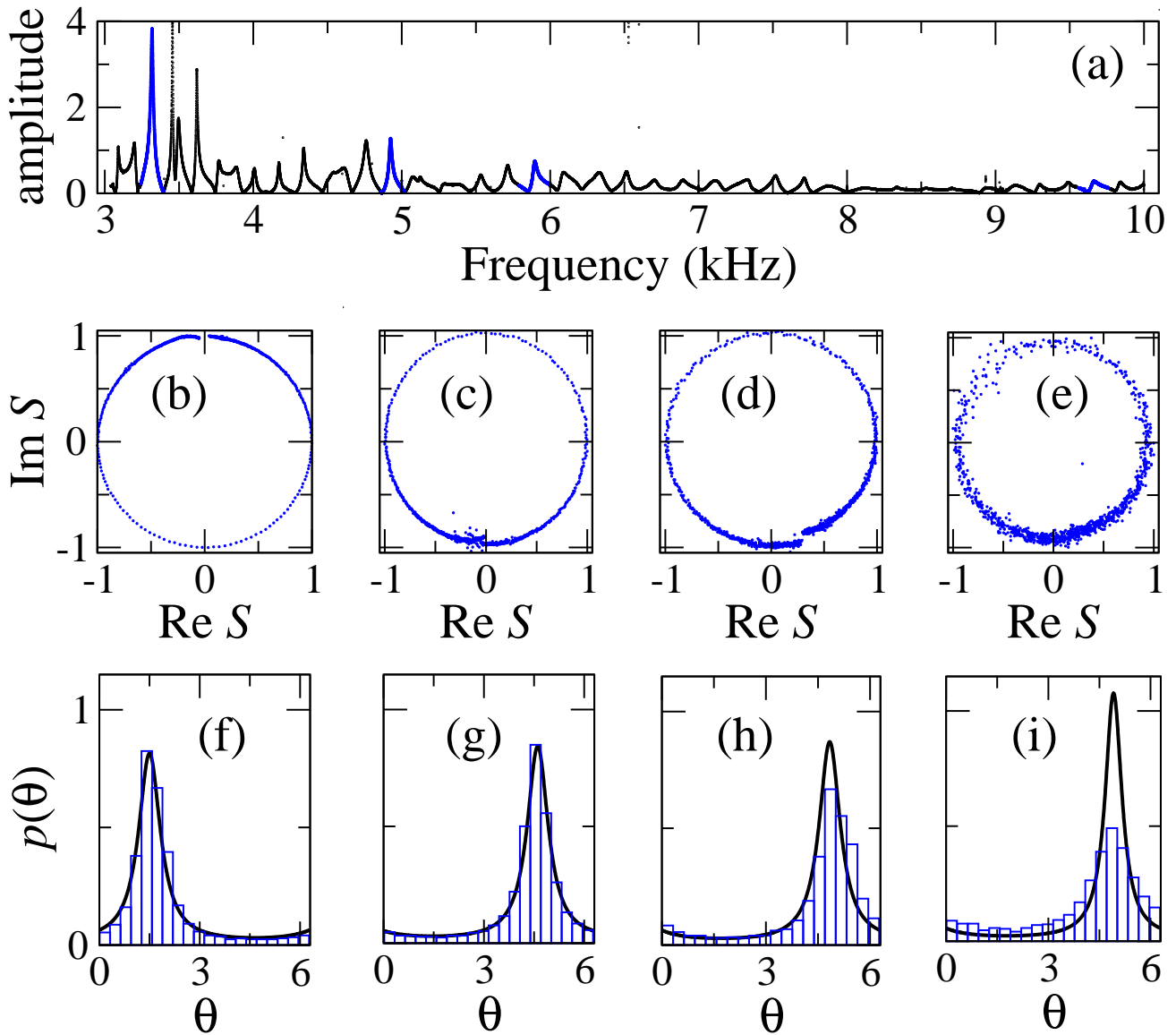}
\caption{(Color online) (a) measured amplitude (in arbitrary units) as a function of the frequency. 
The analyzed resonances (b), (c), (d) and (e), and their respective histograms (f), (g), (h), and (i), are taken in the intervals [3244.0, 3399.0] Hz, [4859.8, 5019.8] Hz, [5792.0, 5998.0] Hz, and [9560.0, 9760.0] Hz, respectively. 
The continuous line in the lower panels is the Poisson kernel, Eq.~(\ref{eq:KPoisson}), with $\langle S\rangle$ given by Eq.~(\ref{eq:Sav}).}
\label{fig:differentresonances}
\end{figure*}
\end{widetext}

{Now the robustness of the expression of the optical matrix, with respect to disorder, losses and noise will be addressed. 
The effect of the disorder in Eq.~(\ref{eq:Sav}) is studied numerically varying randomly
the depth of the groove with a uniform distribution of width $\delta$.} 
That is, the height $h_2$ of body 2 is varied according to $h_2=h+\epsilon(D-h)$, where $\epsilon$ is uniformly distributed in the interval $[0,\,\delta]$. Thus $\delta$ quantifies the disorder strength and $\delta=0$ corresponds to the crystalline structure. 
{In Fig.~\ref{fig:disorder1} the last 15 iterations of 1000 are plotted for different values of the disorder between 10\% and 90\%. 
As it can be seen there, the band structure is preserved for low values of the disorder strength whereas for high disorder the band structure disappears. 
The prediction of the optical matrix agree with the maximum of the distribution for values of the wavenumber in the middle of the first band whereas in the gap and in the second band high deviations are visible.}
%In the upper panel of Fig.~\ref{fig:map} the in frequency is observed for the last 15 iterations of 1000. 
%The allowed bands are clearly observable (the first band and a part of the second one are shown) since the iterations cover the full interval between 0 and $2\pi$ for the phase in a non uniform way. 
%It can be also observed in the forbidden bands, only one is shown, that the iterations reach a single fixed point. 
%The solution in the allowed bands, given by Eq.~(\ref{eq:Sav}), is also plotted. 
%This solution corresponds to the maximum of the distribution around which $\theta_N$ is distributed. 
%In the left lower panel of Fig.~\ref{fig:map} the phase $\theta_N$, for $N=100$, shows the resonances as a function of the frequency in an allowed band; resonances do not appear in the forbidden band. 
{The resulting distributions, for different values of the disorder strength, are shown also in Fig.~\ref{fig:disorder1}. %where we observe that the gaps tend to disappear as disorder increases. 
The distributions of the phase agree with Poisson's kernel, even for very high values of the disorder strength.}
{This result is relevant since it shows the universality of the fluctuations against disorder.}
The used $\langle S\rangle$ was obtained averaging the values of $r_{\rm{n}}$ and $t_{\rm{n}}$ from the disorder. 
%Intermediate values of the disorder (not shown) also agree with the theoretical predictions.
{Some results about the effect of the losses in the optical matrix prediction 
can be obtained directly from the experiment since absorption is always present. 
Contrary to the case in chaotic systems~\cite{Kuhletal,DoCarmoDeAguiar}, in which a generalization of Poisson's kernel appears, in the beam worked here the absorption only gives a shift that can be taken into account in the normalization as in Refs.~\cite{BaezEtAl,Martinez-ArguelloMartinez-MaresCobian-SuarezBaezMendez-Sanchez,Martinez-ArguelloMendez-SanchezMartinez-Mares}.
In Fig.~\ref{fig:differentresonances}(a) the measured amplitude, within the first passband, as a function of the frequency, is given. 
As it can be seen in panels (f) and (g) of the same figure, regardless of the location of the resonance within the allowed band, the agreement between Poisson’s kernel, Eq.~(\ref{eq:KPoisson}), and the experimental results is excellent. 
As it can be seen roughly in panel (a) of this figure, as frequency increases the signal becomes smaller. 
This is evidenced in panels (b), (c), (d) and (e), in which the measured $S$ is plotted for the resonances highlighted in panel (a), with the same order. 
On approaching the forbidden band the signal becomes weak and noisy, panels (d) and (e). Then $S$ does not lie in a circle anymore but it lies in a ring. 
This effect produces a diminishing of the optical matrix, $\langle S \rangle$, that tend to flatten the distribution, panels (h) and (i), in a similar way that in the disordered case.}

Concluding remarks and outlook. The good agreement observed between the theory 
and the experiment represents the validation of the analytical 
expression of the optical matrix $\langle S\rangle$ through the invariant density of the phase of the $S$-matrix, the Poisson kernel, that results from a non trivial relation between coherent transport and deterministic maps~\cite{VidomEPJST}. 
The detected torsional waves outside the locally periodic system, a square cross-section beam with $100$ notches, correspond to the $1\times 1$ scattering matrix $S$ once they are excited also outside. 
The measured distribution of the phase of $S$ of a single 
resonance agrees with Poisson's kernel using the theoretical prediction given in Eq.~(\ref{eq:Sav}); the optical matrix given in that equation depends only on (i) the $S$-matrix composition rule and (ii) the reflection and transmission coefficients of a single unit cell. 
Thus the prediction for the optical matrix is quite general, 
with details depending on the constituents of the particular system, and Eq.~(\ref{eq:Sav})
can be applied to a plethora of wave systems of different 
nature just having the reflection and transmission of a single scatterer. 
Several applications are expected in different areas since there are many realizations of a semi-infinite one-dimensional periodic system composed by two media, {\em i.e.} a photonic crystal, a superlattice, a layered media in geology, etc. 
In fact data of the microwave ring, used to measure the 
Hofstadter's butterfly and transmission through a locally periodic are 
available~\cite{Kuhletal,Luna-AcostaEtAl}. 
The results are also valid for compressional waves~\cite{MoralesFloresGutierrezMendez-Sanchez}. 
Applications in the terahertz~\cite{Dhillonetal} and in the optical~\cite{TribelskyFlachMirishnichenkoGorbachKivsha,PenaGirschikLibischRotterChabanov,ChengMaYepezGenackMello} regimes seem also 
possible since frequency duplicators/dividers can be used to measure the phase.
Another system in which the formalism can be applied is a 1D 
tight-binding chain~\cite{Kittel}. This model is ubiquitous in condensed 
matter~\cite{Izrailev}, material science and chemistry and has several 
applications since the transmission and 
reflection coefficients can also be obtained~\cite{PastawskiEtAlEPL}.
In fact this model has realizations in chains of dielectric 
scatterers~\cite{Franco-VillafaneEtAl,PoliEtAl}, in the evolution of 
excitations in molecular chains and in molecular rings (using nuclear magnetic 
resonance)~\cite{MadiEtAl,Horacio2}.
Microwave billiards~\cite{Mendez-SanchezEtAl2003,Schanze,Hemmadyetal,Kuhletal} and 
graphs~\cite{LawniczakBauchHulSirko} are also well suited to perform different 
tests since the applicability of the equation to cells with more complex 
scatterers, as in a chain of 
cavities~\cite{DittrichEtAl,Luna-AcostaMendez-BermudezIzrailev} is also 
possible. These kind of chains have been constructed with microwave 
billiards~\cite{DembowskiEtAl} and can also be constructed with thin 
plates~\cite{Flores-OlmedoEtAl,GerardinLaurentDerodePradaAubry}.
The Heidelberg approach~\cite{Weidenmuller}, a methodology in which the 
scattering matrix is built in terms of a Hamiltonian and the couplings to the 
exterior, is optimal for applications of Eq.~(\ref{eq:Sav}). 
In fact, results of the prompt response could be possible for nuclear systems, in particular for neutron scattering, which only see the nuclei but not the electrons, would see 
--and identify-- the periodic structure. 
Elastic scattering is part of nuclear reactions, and lots of data are available for (differential) cross sections and scattering of protons, electrons, or neutrons of individual nuclei. 
Although phase shifts are not observables, unlike cross sections, 
it is possible to get non univocal phase shifts from cross sections having a 
model (potential). 
Apart of other effects (as absorption, decoherence, PT-symmetric, more dimensions or a higher number of channels) that have to be included, there are many applications that could be worked out. 
A simple example can be thought in astrophysics: Although in supercooled neutron stars the more recent models are in favour of degenerate baryon models forming stellar 
superfluids~\cite{Page}, within the Bardeen–Cooper–Schrieffer theory, 
there are other models that consider crystallization in one dimension and 
pairing in the perpendicular planes~\cite{Takatsuka}.

\begin{acknowledgments}
This work was supported by DGAPA-UNAM (grant IN109318) and by CONACyT (grant CB-2016/285776). 
VDR and AMMA were supported by DGAPA. 
We thank G. B\'aez, A. M\'endez-Berm\'udez, T. H. Seligman, H. Pastawski, T. Papenbrook, and D. Page for useful comments and to ``Centro Internacional de Ciencias A. C.'' for several meetings held there and the given facilities for the laboratory.
\end{acknowledgments}

\end{document}